\documentclass[sigconf]{acmart}
\AtBeginDocument{%
  }

\setcopyright{none}
\acmYear{}
 \acmDOI{}
 \acmISBN{}
\acmConference[FAccTRec@RecSys]{Workshop on Socially Responsible Recommender Systems}{2025}

\begin{document}

\title{Cascade! Human in the loop shortcomings can increase the risk of failures in recommender systems}


\author{Wm. Matthew Kennedy}
\affiliation{%
  \institution{Oxford Internet Institute}
  \institution{University of Oxford}
  \city{Oxford}
  \country{United Kingdom}}
\email{matt.kennedy@oii.ox.ac.uk}

\author{Nishanshi Shukla}
\affiliation{%
  \institution{Western Governors University}
  \city{Millcreek}
  \state{UT}
  \country{United States}}
  \email{nishanshi.shukla@gmail.com}

\author{Cigdem Patlak}
\affiliation{%
 \institution{Independent}
 \city{Irvine}
 \state{CA}
 \country{United States}}
 \email{cigdemp@gmail.com}

\author{Blake Chambers}
\affiliation{%
  \institution{Independent}
  \city{Boston}
  \state{MA}
  \country{United States}}
  \email{blakejwc@alum.mit.edu}

\author{Theodora Skeadas}
\affiliation{%
  \institution{Humane Intelligence}
  \city{Boston}
  \state{MA}
  \country{United States}}
\email{Theodora@humane-intelligence.org}

\author{Tuesday}
\affiliation{%
  \institution{ARTIFEX Labs}
  \country{United States}}
\email{Tuesday@artifex.fun}

\author{Kingsley Owadara}
\affiliation{%
  \institution{Pan-Africa Center for AI Ethics}
  \city{Lagos}
  \country{Nigeria}}
\email{Kowdara@yahoo.com}

\author{Aayush Dhanotiya}
\affiliation{%
  \institution{Amazon Inc.}
  \city{Seattle}
  \state{WA}
  \country{United States}}
\email{aayush.dhanotia@gmail.com}

\maketitle

\section{Introduction}
Recommender systems are among the most commonly deployed systems today. Systems design approaches to AI-powered recommender systems (that is, systems that are both data-driven and employ computationally expensive algorithms such as neural networks) have done well to urge recommender system developers to follow more intentional data collection, curation, and management procedures \cite{chen2023, delValle2023, stray2024}. So too has the “human-in-the-loop” paradigm been widely adopted (at least nominally), primarily to address the issue of accountability \cite{fu2021, ustalov2022, vanVoorst2024}.

However, in this paper, we take the position that human oversight in recommender system design also entails novel risks that have yet to be fully described. These risks are “codetermined” \cite{weidinger2023sociotechnicalsafetyevaluationgenerative} by the information context in which such systems are often deployed. Furthermore, new knowledge of the shortcomings of “human-in-the-loop” practices to deliver meaningful oversight of other AI systems suggest that they may also be inadequate for achieving socially responsible recommendations. We review how the limitations of human oversight may increase the chances of a specific kind of failure: a “cascade” or “compound” failure. We then briefly explore how the unique dynamics of three common deployment contexts can make humans in the loop more likely to fail in their oversight duties. We then conclude with two recommendations.

\section{The cascade problem}
Among the best known failure modes of recommender systems is the “data cascade” failure. Cascade failures might be thought of as compound failures, in that they are caused by several apparently trivial failures bootstrapping into a larger, more significant failure. These properties make them exceedingly difficult to identify, much less mitigate, in production systems. Likewise, they subject users of that system to a prolonged series of low-value recommendations, which may in fact accelerate the compounding effect of a cascade failure. Cascade failures occur for three reasons primarily: low data quality, poor system design, and insufficient oversight (including human oversight). It is the latter category on which this paper focuses \cite{sambasivan2021}. 

\section{Current human-in-the-loop approaches may be insufficient}

The scientific literature on human oversight of recommendation systems is growing but not yet mature. Where human-in-the-loop approaches have been discussed or applied in the context of improving recommender systems, it has usually taken the form of human evaluation of the information retrieved via user searches or the information quality of the search space itself. As a result, the literature has tended to focus on search-based recommender systems \cite{ustalov2022} or as a proposed measure to resolve longstanding technical challenges to recommender system design – for instance the “long tail” problem \cite{fu2021} that requires systems to produce recommendations for items or content that itself is not highly subscribed \cite{zhao2023}.

With time, however, we are beginning to understand that “human in the loop” practices vary widely in their configuration and indeed in their quality \cite{tsagas2024}. This variance is often a result of human component failures. Human in the loop failures take many forms that are consequential to recommender systems. For instance, recent research from the field of algorithmic decision-making has demonstrated that humans in “in the loop” governance functions provided “correct” oversight only about half the time - lapses primarily caused by human motivation to ensure compliance with their organization’s goals rather than responsible AI principles \cite{gaudeul2025}. 

Considering that humans in the loop in recommender systems are involved “upstream” in data quality evaluation functions, humans in the loop may be inadequate to regularly perform such data quality duties. AI-powered recommender systems may pose special challenges. For instance, humans tasked with evaluating the data quality or information retrieval components of a conversational recommender system that employs LLMs to elicit user preferences will encounter highly context-specific natural language. Their assessments may introduce new errors, inadvertently pushing the system into lower value areas of its search space, or causing the system to make low-quality associations (Squires 2006). If these kinds of degradations of the information context are not identified and mitigated quickly, they can cause recommender systems to deliver recommendations that may at first appear appropriate but that eventually lead the system (and thence the user) down the path of compounding diminishing returns \cite{wachter2024a}. In the worst case, such failures can lead to model collapse \cite{shumailov2024}.

\section{Examples}

To concretize this conceptual exploration, we briefly review three different 'information contexts' in which recommender systems are often deployed and where incorrect solutions to long-tail problems are likely to have more substantial consequences: education, social media, and e-commerce.

\subsection{Example 1: education}

As education technology application developers embrace AI in their offerings, several claim to have developed 'personalized' or 'adaptive' learning solutions that renovate previous intelligent tutoring systems (ITS) to respond to individual student needs \cite{holmes2018technology}. However, many edtech researchers are critical of these claims, pointing out that 'personalization' is rarely truly personalized and should rather be thought of as 'pre-programmed' \cite{jurenka2024responsibledevelopmentgenerativeai, holmes2018technology}. They also warn that classroom use of such systems can subtly steer what kinds of learning content a student encounters, affecting student mastery, advancement, and therefore educational attainment \cite{unesco2023}. The stakes in these environments are high, and the margin for error is small. A misstep in what gets recommended or what gets withheld can shape a student’s academic path in ways that are difficult to detect or reverse \cite{KennedyForthcoming}. Furthermore, human oversight of such systems is difficult as application developers rarely provide real-time monitoring, and teacher interaction usually only comes after the fact, through dashboards, summary reports, or assessment. 

\subsection{Case 2: content recommender systems in social media}

Recommender systems can facilitate the sharing of content from conspiracy-oriented channels, hateful content, and divisive content \cite{buntain2021} which fuel misinformation, political division, and radicalization. Moderation algorithms make millions of content and account removal decisions daily, but many of these decisions are incorrect, informed by poor language and context understanding and therefore improper prioritization, a consequence of under-resourced human oversight. In some cases, this underresourcedness leads to unbalanced deployment of algorithmic content moderation tools. For example, a human rights due diligence of Meta’s impacts in Israel and Palestine by Business for Social Responsibility found that “proactive detection rates of potentially violating Arabic content were significantly higher than proactive detection rates of potentially violating Hebrew content...[likely because] there was an Arabic hostile speech classifier but not a Hebrew hostile speech classifier” \cite{bsr2022}. Additionally, human oversight resources are concentrated on high-value accounts, a concept referred to as 'analog privilege' \cite{levesque2024}. 

\subsection{Case 3: recommender systems in e-commerce}

Recommender systems are widely deployed to e-commerce platforms and heavily influence user perceptions of the goods or services available for purchase and, in some cases, also producing “personalization” in pricing, though this practice is increasingly prohibited \cite{Seele2019-SEEMTE, gautier2020}. In either case, recommender systems in e-commerce are critical parts of a platform’s sales funnel, aiding in keeping users engaged and moving towards a transaction by surfacing useful results to queries and in helping sellers market their products to likely buyers \cite{sorokina2016}. Of course, such systems perennially encounter the long tail problem, and have been criticized for appearing to recommend already items that are already popular, making it harder for smaller sellers or more niche, ethical, or sustainable options to surface, especially on very large platforms. These systems also serve the interests of platform providers themselves, and precisely how proprietary search-based recommender systems retrieve potential recommendations is not always clear to consumers, sellers, or regulators. Many of the largest e-commerce platform providers have been accused of deploying recommender systems that unfairly shape commercial information contexts to favor their own platform or the goods that provide them with greater commercial benefits instead of items that may actually better match user searches and preferences \cite{strauss2024}. Over time, this can subtly shape market perception and consumer behavior. More germane to this paper, this places humans in the loop in a sustained dilemma where they must choose which values–their organization’s or those that comprise responsible AI–to uphold when they observe conflicts. There is now emerging evidence to suggest that, when faced with such a dilemma, humans in the loop are just as likely to align to one as with the other \cite{gaudeul2025}.

\section{Recommendations for the field}

Clearly, there is more work to be done mapping out this particular area of risk. We hope, however, that our paper establishes the importance of doing this work. To that end, we conclude with two of our own recommendations for the field of socially responsible recommender system design. Firstly, we urge recommender system designers to consider more deeply the role they are expecting humans to play in delivering high-value recommendations in all kinds of recommender systems. Further and more intentional review of the human components of these systems is critical in order to avoid their becoming “moral crumple zones” \cite{elish2019}. Second, we urge systems designers to investigate the limitations of humans in evaluation functions for recommenders operating over very large search spaces that require opaque ML/AI algorithms to process. Given the critical role humans in the loop have been expected to play in ensuring data and search quality (and therefore recommendation quality), greater attention should be paid to the potential shortcomings and misallocations of human oversight capabilities.

\bibliographystyle{ACM-Reference-Format}
\bibliography{facctrecbib}

\end{document}